# Exploring the Big Bang with femtoscopy

MÁTÉ CSANÁD

ELTE Eötvös Loránd University, Dept. of Atomic Physics, Pázmány P. s. 1/a, 1117 Budapest, Hungary.

Email: csanad@elte.hu

**Abstract**

Exploring the fundamental constituents of the matter around us and in the Universe, as well as their interactions, is among the premier goals of physics. Investigating ultrarelativistic collisions in particle accelerators has delivered answers to these questions many times in the past decades. In this paper we focus on the research aimed at recreating the matter that filled the Universe in the first microsecond after the Big Bang – but this time in collisions of heavy ions. In particular we discuss the technique called femtoscopy, which provides us a tool to understand the space-time structure of particle creation in heavy-ion collisions. We utilize Lévy-stable distributions to investigate this structure and explore its dependence on particle momentum and collision energy.

**Introduction**

The strong interaction is the fundamental interaction of Nature responsible for confining quarks and gluons, i.e. the constituents of protons and neutrons, as well as for most of the mass of the observable Universe. Along a series of discoveries (Adams 2005, Adcox 2005, Back 2005, Arsene 2005), it turned out in the last decades that in ultrarelativistic collisions of heavy nuclei, the so-called strongly interacting Quark Gluon Plasma (sQGP) is created, which is governed by the strong interaction. This matter, the sQGP filled the Universe for the first microsecond of its existence. In this medium, the charges of the strong interaction, i.e., coloured objects are deconfined. At large collision energies, such as the ones available at the Relativistic Heavy Ion Collider (RHIC) or the Large Hadron Collider (LHC), the sQGP continuously transitions to form a hadronic matter, where colour is confined into colourless hadrons: baryons and mesons. This matter then expands, and particle detectors built around the collision point register its constituents. At smaller collision energies, a first order phase transition from deconfined to hadronic matter is expected, albeit not confirmed experimentally. Between the two collision energy regions, a Critical Endpoint (CEP) of the phase diagram of the strongly interacting matter may be present, as illustrated in Figure 1. This figure shows the phases of strongly interacting matter: at low baryon densities (quantifying the energy related to the baryon excess) and temperatures hadronic matter is present, where quarks are confined into hadrons. At large densities or temperatures, colour may be deconfined, and the sQGP or other, exotic forms of matter (e.g., in neutron stars) may be present.

One of the most important goals of today's high-energy heavy ion physics is to confirm (or experimentally rule out) the existence of the CEP, and if it exists, to characterize it. At various particle accelerator facilities, beam energy scan programs were initiated to study this phase diagram. To do so, femtoscopic (momentum) correlations can be utilized to gain information about the space-time geometry of particle emission (Ledniczky 2001). In this paper, the utilization of Lévy-stable distributions is discussed, generalizing the conventional Gaussian analysis for the source geometry (Csörgő 2004). Furthermore, recent measurements performed at various experiments are presented, as well as corresponding phenomenological interpretations that can be utilized to explore the phase diagram of strongly interacting matter.

The paper is divided into six sections. In the first section, the broader field is introduced. In the second section, the HBT effect is discussed, and in the third the subfield femtoscopy is detailed. Then in the fourth section the utilization of Lévy-stable sources is discussed, while in the fifth section, experimental results and interpretations are mentioned. Finally, the sixth section findings are briefly summarized.

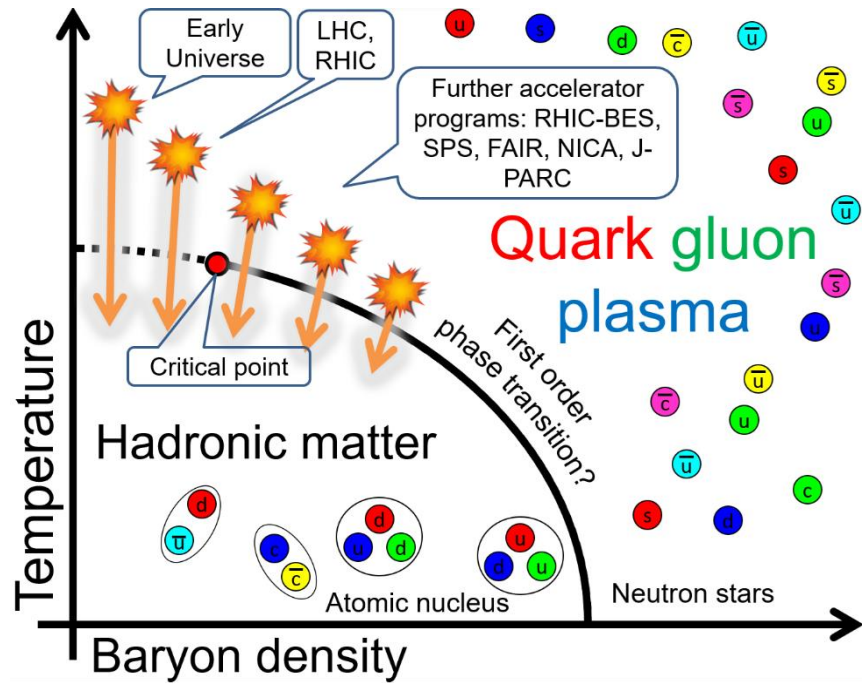

*Figure 1: Phases of the strongly interacting matter, as a function of temperature and baryon density. Arrows illustrate that lower energy collider experiments probe the phase diagram at larger barion densities and smaller temperatures.*

**The HBT effect**

Before going through the methodology of measuring and interpreting femtoscopic correlations in high-energy physics, let us mention that the origins of this field go back to R. Hanbury Brown and R. Q. Twiss, who investigated stars with optical and radio telescopes, at Jordell Bank and later at the Narrabri Observatory (Hanbury Brown, 1956). They discovered an unexplained correlation between signals of two telescopes, interpreted as coincidental photon detections happening "too frequently" at the two telescopes, as compared to the expected frequency of random coincidences. They furthermore observed that this correlation (excess number of coincidences) is decreased when increasing the distance between the two telescopes. Hanbury Brown later wrote in his autobiography (Hanbury Brown 1991):

> "In fact, to a surprising number of people the idea that the arrival of photons at two separated detectors can ever be correlated was not only heretical but patently absurd, and they told us so in no uncertain terms, in person, by letter, in print, and by publishing the results of laboratory experiments, which claimed to show that we were wrong."

Indeed, while these experiments were repeated in tabletop experiments as well, it took several decades to unravel the physics and mathematics behind this phenomenon, now called HBT effect. There are two simple ways of understanding it. First is (based on Baym 1998) when one regards the optical or radio signal as spherical waves being emitted from a point $a$ within the source, with the amplitude

$$A_a(r) = \frac{1}{|r-r_a|} \alpha e^{ik|r-r_a|+i\Phi_a},$$

where $r$ is the spatial point we investigate the wave at, $r_a$ is the origin of the wave, $\alpha$ is its strength, $k$ its wave-number, and $\Phi_a$ is its phase, assumed to be a random variable with a uniform distribution in case of thermal emission. A similar amplitude can be expected from another point of the source, let us denote that one with a similar expression, but with index $b$ and strength $\beta$. Let us assume that the source comprises only these two points, denoted by $a$ and $b$. Then in any point in space, the wave amplitude from the source can be written as

$$A(r) = A_a(r) + A_b(r).$$

At the detectors, a sum of the waves emitted from both points of the source is detected. The observed intensity in detector $A$ is then the square of the amplitude:

$$I_A = |A(r_A)|^2 = \frac{1}{L^2}\left(|\alpha|^2 + |\beta|^2 + \alpha^*\beta e^{ik(r_{bA}-r_{aA})+i(\Phi_b-\Phi_a)} + c.c.\right),$$

where $r_A$ is the location of detector $A$, $L$ is the average distance between the source points and the detectors, while $r_{bA}$ is the distance of source point $b$ and detector $A$, similarly for $r_{aA}$, and c.c. denotes the complex conjugate of the preceding term. The observed signal is then this intensity, averaged over a sufficiently large time interval, where phase information is cancelled if the source is chaotic:

$$\langle I_A \rangle = \langle I_B \rangle = \frac{1}{L^2}(|\alpha|^2 + |\beta|^2).$$

The joint (or coincident) intensity is however the product of intensities, where the phase information is retained even after averaging, resulting in an excess term as compared to the product of time-averaged intensities:

$$\langle I_A I_B \rangle = \langle I_A \rangle \langle I_B \rangle \left(1 + \frac{1}{2}\cos k(r_{aA} - r_{bA} + r_{aB} - r_{bB})\right), \text{ and}$$

$$C_{AB} = \frac{\langle I_A I_B \rangle}{\langle I_A \rangle \langle I_B \rangle} \approx 1 + \frac{1}{2}\cos\frac{kRd}{L},$$

where the last approximation can be confirmed by expanding the sum of the distances in terms of $d/L$ and $R/L$, where $d$ is the distance of the detectors and $R$ is the distance of the investigated source points. The factor of $1/2$ in front of the cosine is specific to this configuration of two source points, the general result is the excess term, decreasing with increasing detector distance or source size. In a general source, an origin-dependent source strength $\alpha(r_s)$ would appear, and the sum would go over all possible $r_s$ values (in fact it would an integral) instead of the $A_a(r) + A_b(r)$ sum written above.

A second possible understanding of the HBT effect is when one considers the detection of two quantum-mechanical particles stemming from sources $a$ and $b$, in detectors $A$ and $B$. If we regard the two detection events separately, then the corresponding wave-functions are

(assuming plane-waves here for simplicity, but the calculation would work equivalently for spherical waves as well):

$\Psi_B^b = e^{ikr_{bB}+i\Phi_b}$ and $\Psi_A^a = e^{ikr_{aA}+i\Phi_a}$.

If these particles are identical bosons (as are photons), then the two-particle wave-function must be symmetric w.r.t. particle exchange (c.f. Bose-Einstein statistics), and the detection events are also entangled, expressible with the wave-function

$\Psi_{AB} = \frac{1}{\sqrt{2}}\left(\Psi_A^a \Psi_B^b + \Psi_B^a \Psi_A^b\right)$.

Then the single and joint detection probabilities are the modulus squares of the above wave-functions for a single particle and the pair, respectively. In case of a setting similar to the one discussed in the classical derivation, the result in this case is:

$C_{AB} = \frac{P(A,B)}{P(A)P(B)} = 1 + \cos k(r_{aA} - r_{bA} + r_{aB} - r_{bB}) \approx 1 + \cos k\frac{Rd}{L}$.

This second derivation can be extended to fermions as well, in which case a similar anti-correlation is observed. Furthermore, for an extended source, a source weight could appear as a prefactor in $\Psi$ and an integral over $r$ would have to be performed. Figure 2 shows example data from the measurement of Sirius by Hanbury Brown and Twiss, along with a calculation of such an extended source, reproduced from Hanbury Brown, 1956.

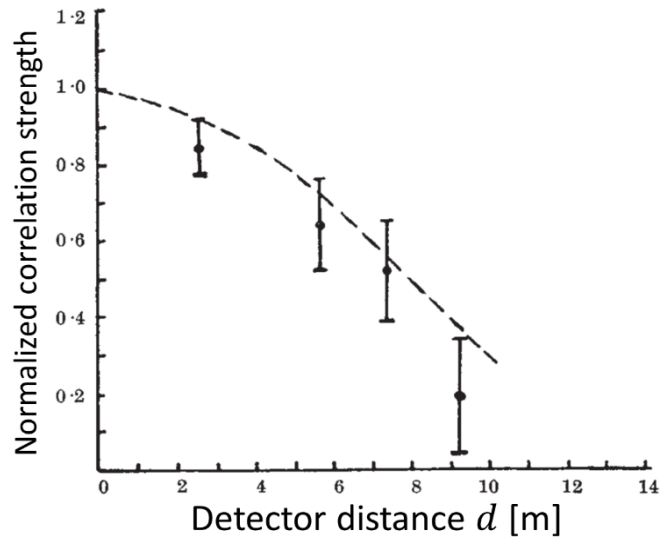

*Figure 2: Correlation data from Sirius and a theoretical curve based on the calculation of a star of angular size 0.0063'', reproduced from Hanbury Brown 1956.*

**Femtoscopy**

Independently of the HBT-discovery, Goldhaber and collaborators observed a similar correlation for pion pairs (Goldhaber, 1959), detected in high-energy proton-proton collisions. A clear understanding of the background of the two related phenomena is owed to papers by Glauber, Baym and others (Glauber 1963, Baym 1998). The main result, applicable today in high-energy physics is that the two-particle momentum correlation function is connected to the particle creation probability density. This can be understood based on the connection of the single-particle momentum distribution $N_1(p)$ expressed as

$$N_1(p) = \int S(r,p)|\Psi_p(r)|^2 dr,$$

where $S(r,p)$ is the particle creation probability density (also called single-particle source or simply source) at the space-time point $r$ and momentum $p$ (in other words, the infinitesimal probability of creating a particle with momentum $p$ in point $r$, which would be a sum of two Dirac delta distributions in the above two-source example), $\Psi_p(r)$ is the single-particle wave-function of a particle with momentum $p$, and the integral runs over the entire space-time. Similarly, the two-particle momentum distribution $N_2(p_1, p_2)$ can be expressed as

$$N_2(p_1, p_2) = \int S(r_1, p_1) S(r_2, p_2) |\Psi_{p_1, p_2}(r_1, r_2)|^2 dr_1 dr_2, \text{ where}$$

$$\Psi_2(r_1, r_2) = \frac{1}{\sqrt{2}} \left( \Psi_{p_1}(r_1) \Psi_{p_2}(r_2) + \Psi_{p_2}(r_1) \Psi_{p_1}(r_2) \right)$$

is the two-particle symmetrized wave-function (for identical bosons). Assuming plane-waves (i.e., neglecting interactions), defining $q = p_1 - p_2$, $K = (p_1 + p_2)/2$ and utilizing the approximation where $q \ll K$, one obtains for the correlation function of identical bosons

$$C(q,K) = \frac{N_2(q+K, q-K)}{N_1(q+K) N_1(q-K)} \approx 1 + \frac{|\int dr S(r,K) e^{iqr}|^2}{|\int dr S(r,K)|^2} = 1 + \frac{|\tilde{S}(q,K)|^2}{|\tilde{S}(0,K)|^2},$$

where the tilde (in $\tilde{S}$) denotes the Fourier transform in the first variable. If one introduces the spatial correlation function as

$$D(r,K) = \int S\left(\rho + \frac{r}{2}, K\right) S\left(\rho - \frac{r}{2}, K\right) d\rho,$$

and assumes it to be normalized to unity, then one obtains

$$C(q,K) = 1 + \int dr D(r,K) e^{iqr} = 1 + \tilde{D}(q,K).$$

The main result here is that the momentum correlation function is directly and simply connected to the spatial (pair-) distribution. In femtoscopy, one often assumes a parametric shape of for the $r$-dependence of $S$ or $D$, and then the momentum dependence is expressible through the dependence of the parameters on $K$. For example, if the source is a Gaussian:

$$S(r,K) \sim \exp\left[-\frac{r^2}{2R(K)^2}\right], \text{ or equivalently } D(r,K) \sim \exp\left[-\frac{r^2}{R(K)^2}\right], \text{ then}$$

$$C(q,K) = 1 + \exp[-q^2 R(K)^2],$$

hence a source of some given width $R$ creates correlations of inverse width $1/R$. This is illustrated by Figure 3; and this width is then also called HBT radius (after the HBT effect).

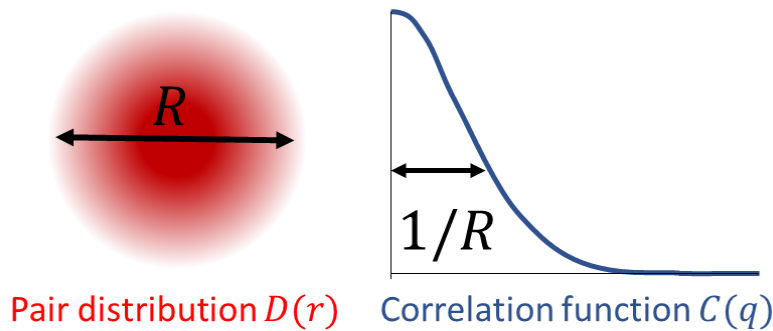

Figure 3: A source of width R creating correlations of with 1/R.

This simple picture neglects several phenomena, such as final-state interactions (FSI), secondary particle production and multiparticle correlations. For details about these, see for example References Adare 2018 and Nagy 2023. We only note here that in case of secondary particle production from decays of long-lived resonances, the correlation strength decreases (Adare 2018), and the correlation function (for the same Gaussian source as above) becomes

$$C(q,K) = 1 + \lambda(K) \cdot \exp[-q^2 R(K)^2],$$

where $\lambda(K)$ is the (momentum-dependent) correlation strength, which can be interpreted as the squared fraction of the primordially produced particles (Adare 2018). Let us note at this point that in experimental analyses, usually the dependence on $m_T = \sqrt{m^2 + K^2}$ (where $m$ is the identified particle's mass) is investigated instead of the dependence on $K$.

**Lévy-stable sources**

Due to the central limit theorem and thermodynamics leading to Gaussians, the natural assumption for the source shape is that of a Gaussian, and this assumption has been explored in a multitude of experimental and phenomenological investigations (Lisa 2005). In this paper we refrain from discussing those and focus on a different avenue of research instead. As mentioned in several papers (Csörgő 2004, Csanád 2007, Adare 2018), measurements (see e.g. Adler 2007 or references in Adare 2018) suggest phenomena beyond the Gaussian distribution. A straightforward generalization of the central limit theorem for random variables without a finite variance or mean leads to the general class of Lévy-stable distributions (Csörgő 2004). Out of these we focus here on symmetric Lévy distributions, defined as

$$\mathcal{L}(r;\alpha,R) = (2\pi)^{-3} \int dq e^{iqr} e^{-\frac{1}{2}|qR|^\alpha},$$

where $r$ is the (three-dimensional) variable of the distribution, Lévy-index $\alpha$ and Lévy-scale $R$ are its parameters, $q$ is an integration variable, and $qr$ is the scalar product of $q$ and $r$. The distributions are stable for $0 < \alpha \leq 2$, meaning that a sum of random variables with such distribution converges to a random variable with the same distribution. In case of $\alpha < 2$ and three dimensions, these distributions exhibit a power-law tail with exponent $-(1+\alpha)$. For $\alpha = 2$, the Gaussian distribution is retained, while $\alpha = 1$ one obtains the Cauchy (Lorentz) distribution. Note also that the Lévy-scale $R$ can also be called the HBT-radius, but it has to be kept in mind that it is conceptually a different observable as the one extracted from Gaussian measurements (except if $\alpha$ is close to 2).

If the single-particle source is of the above discussed Lévy-stable shape, then the pair-source is also Lévy-stable with the same exponent, but a modified width:

$$S(r) = \mathcal{L}(r;\alpha,R) \Rightarrow D(r) = \mathcal{L}(r;\alpha,2^{1/\alpha}R),$$

where we dropped the dependence on mean momentum $K$ for simplicity.

There may be several reasons for the appearance of these distributions, see a short review in Ref. (Csanád 2024). One example is anomalous diffusion (Csanád 2007), as illustrated in Figure 4, where it is also apparent that in case of anomalous diffusion (also called Lévy flight) the appearance of large step sizes is much more frequent than in normal diffusion, creating a heavy tail in the resulting distribution.

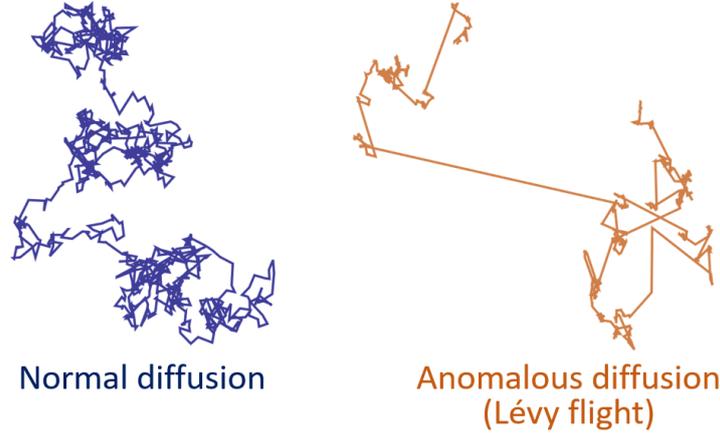

Normal diffusion        Anomalous diffusion
                        (Lévy flight)

*Figure 4: Examples for normal and anomalous diffusion in two dimensions. Images reproduced based on Wikipedia, 2024.*

Another example for the reasons of appearance of Lévy-stable distributions may be a second-order phase transition in the vicinity of the CEP, as detailed in Csörgő 2006. It has been conjectured that at the critical point, the above introduced Lévy-exponent $\alpha$ is identical to the critical correlation exponent $\eta$, hence drops to a much smaller value than 2, even down to values near 0.5 (Csörgő 2006, Csanád 2024).

**Experimental results and interpretations**

As detailed above, femtoscopy can be utilized to explore the probability density distribution of particle creation from the sQGP. The most abundant particles are pions; hence their correlations can be measured most precisely (in terms of statistical uncertainty). The considerations in the previous section motivate the accurate measurement of the Lévy-exponent $\alpha$, but also measuring $R$ with the correct source assumption is of crucial importance (Csanád 2024). In this section we discuss several recent experimental results on these parameters.

The detailed centrality and average transverse mass dependence of these two-pion Lévy HBT parameters have been recently measured in 5.02 TeV Pb+Pb collisions with the CMS experiment at the LHC (Tumasyan, 2024). Here no particle identification was applied, instead all detected particles were assumed to be pions. This reduces the correlation strength but does not change its shape, as non-pions do not contribute to the measured correlations. This analysis found qualitatively similar trends of the $m_T$-dependence of Lévy-parameters as the first measurement of Lévy-parameters in heavy-ion collisions, measured in 200 GeV Au+Au collisions by PHENIX at RHIC (Adare, 2018):

- a nearly constant $\alpha(m_T)$,
- a nearly constant $\lambda(m_T)$ at large $m_T$ (when rescaled by the pion fraction),
- and an $R$ decreasing with $m_T$ as $R \propto \sqrt{m_T}$.

However, measured $\alpha$ values obtained by CMS were found to be around 1.6-2.0, significantly larger than those of PHENIX (where a mean value of $\alpha \approx 1.2$ was found in 0-30% Au+Au collisions). The $\alpha$ and $R$ values measured by CMS are shown in Figures 5 and 6.

A similar analysis in STAR was also performed in 200 GeV Au+Au collisions (Kincses 2024). Similar values, $m_T$-dependence, and centrality-dependence for the Lévy-index $\alpha$, Lévy-scale $R$ and correlation strength $\lambda$ were found as in Adare, 2018. Let us note in particular the decrease of $\lambda(m_T)$ at small $m_T$, attributed to a possible indirect signal of in-medium mass modification of the $\eta'$ meson (Adare, 2018). The trends of the parameters versus $m_T$ are also similar to the above-mentioned CMS analysis (noting the difference in the accessible $m_T$ interval); while the largest difference is that in the STAR measurement, $\alpha$ decreases for central collisions. Results by STAR on $\alpha$ and $R$ are shown in Figures 7 and 8.

Measurements by the NA61/SHINE experiment at the Super Proton Synchrotron (SPS) were reported for Be+Be collisions in Adhikary, 2023 and for Ar+Sc collisions in Pórfy, 2024. These collision systems are initially significantly smaller than those in Au+Au collisions at RHIC and in Pb+Pb collisions at the LHC. The $\alpha$ values measured in Be+Be collisions were found to be around 1.6-2.0, similar to those obtained at the LHC. The ones measured in Ar+Sc collisions were found to be around 1.0-1.4, similar to those obtained at RHIC. The origin of this behaviour is yet unexplained. The Lévy-scale $R$ shows a similar trend as the one at RHIC or LHC, but with smaller values, corresponding to the much smaller initial system size. Results from NA61/SHINE on $\alpha$ and $R$ are shown in Figures 9 and 10. Note furthermore that the correlation strength $\lambda$ does not show the characteristic decrease at small $m_T$ observed at RHIC (and mentioned above). This hints at a difference in the occurring physical processes at SPS and at RHIC.

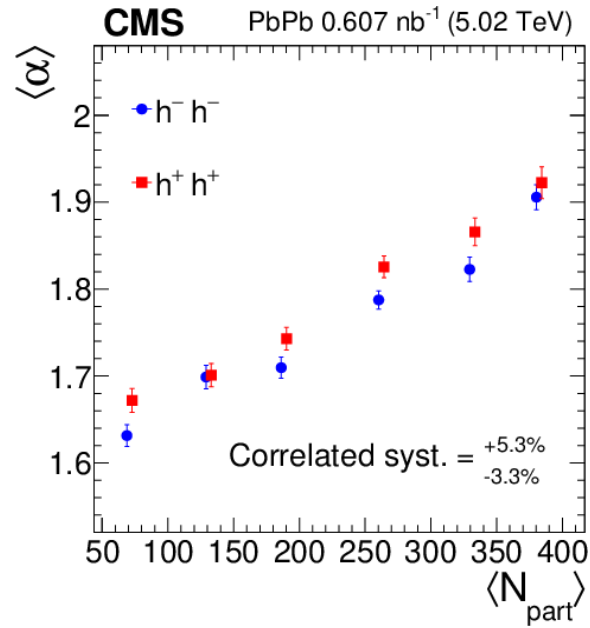

*Figure 5: Lévy-index $\langle\alpha\rangle$ averaged over pair transverse mass, as a function of the number of participant nucleons ($N_{part}$) in the collision, measured by CMS at LHC. Figure reproduced from Tumasyan, 2024.*

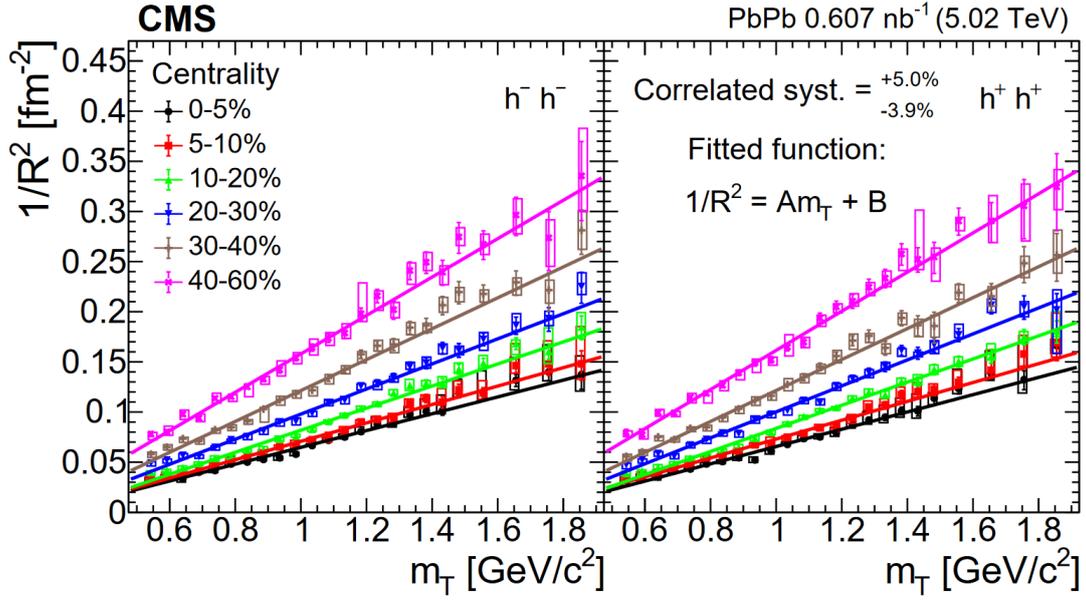

Figure 6: Lévy-scale R as a function of pair transverse mass $m_T$, measured by CMS at LHC. Solid lines show linear fits based on hydrodynamic predictions. Figure reproduced from Tumasyan, 2024.

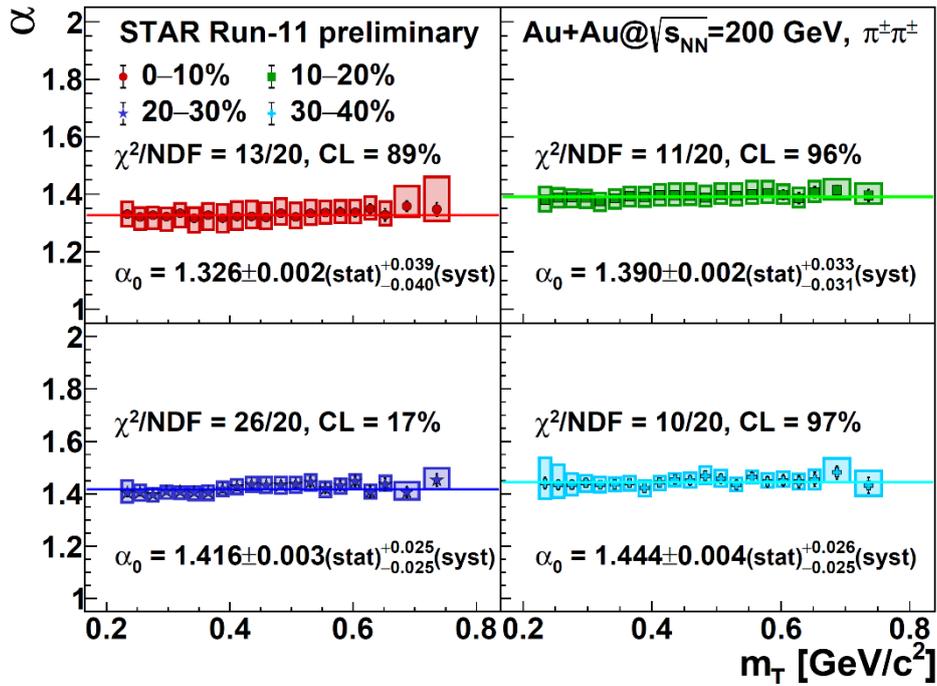

Figure 7: Lévy-index $\alpha$ as a function of pair transverse mass $m_T$, measured by STAR at RHIC. Solid lines show constant fits. Figure reproduced from Kincses, 2024.

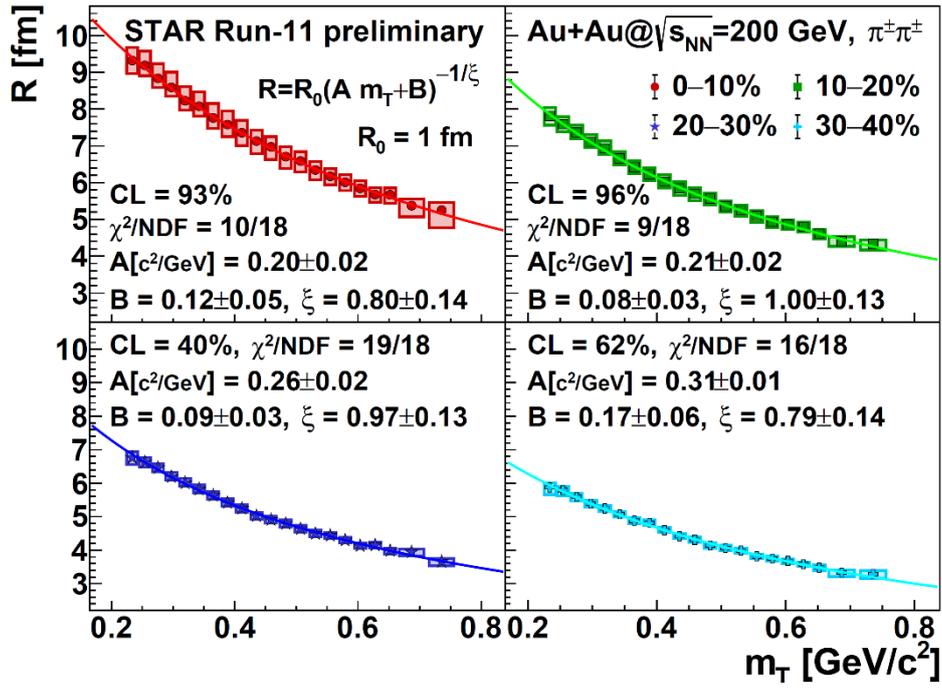

Figure 8: Lévy-scale $R$ as a function of pair transverse mass $m_T$, measured by STAR at RHIC. Solid lines show fits with an empirical curve. Figure reproduced from Kincses, 2024.

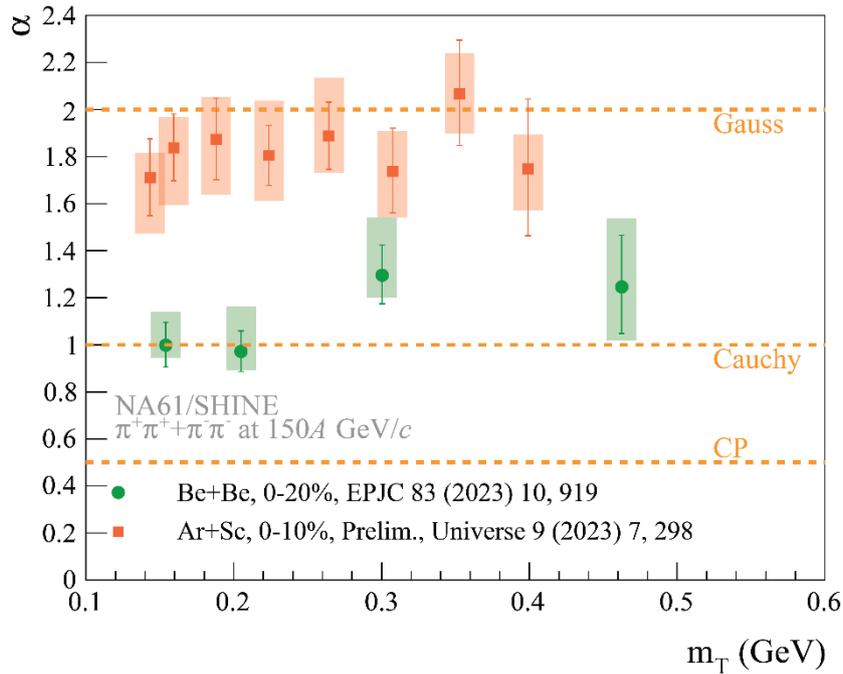

Figure 9: Lévy-index $\alpha$ as a function of pair transverse mass $m_T$, measured by NA61/SHINE at SPS. Also shown are lines for special values of $\alpha$. Figure reproduced from Pórfy, 2024.

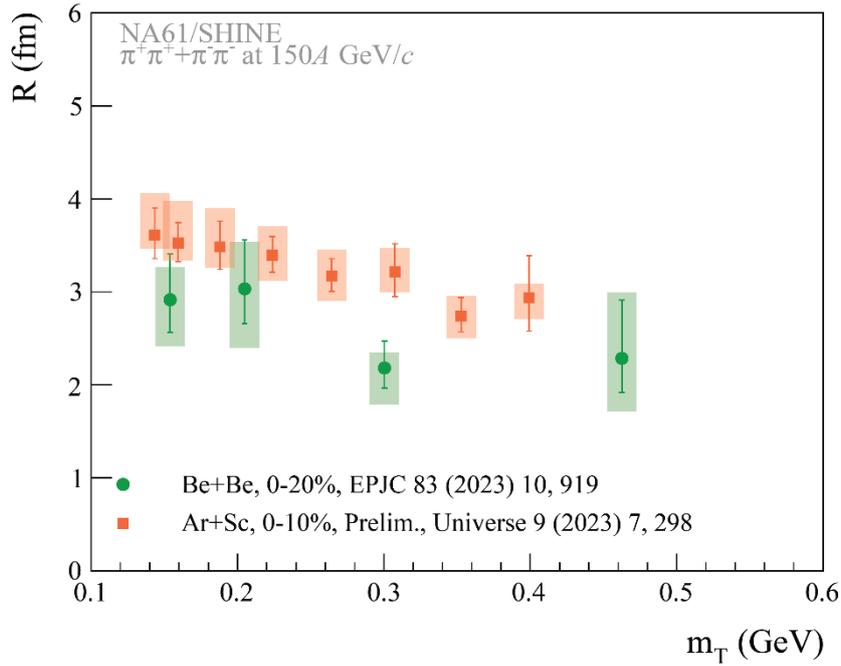

Figure 10: Lévy-scale $R$ as a function of pair transverse mass $m_T$, measured by NA61/SHINE at SPS. Figure reproduced from Pórfy, 2024.

**Summary**

Several phenomenological and experimental results related to femtoscopy were presented above. It was found that Lévy sources are observed at a wide range of collision energies, at SPS, at RHIC, and at the LHC. Multiple physical phenomena may influence the appearance of these sources: two examples discussed above are anomalous diffusion and critical phenomena. At LHC energies, as well as at higher RHIC energies, only the first (anomalous diffusion) may influence experimental correlation functions, as no critical phenomena are expected in this collision energy range. At lower RHIC energies or at the SPS, these may play a role as well. Observations, on the other hand, show varying values of Lévy-exponent $\alpha$: they are around 1.6-2.0 in SPS Be+Be collisions and LHC Pb+Pb collisions, while $\alpha$ values around 1.0-1.4 are found in SPS Ar+Sc collisions and RHIC Au+Au collisions. Furthermore, opposite centrality dependence of $\alpha$ was found at RHIC and at LHC. Hence it is clear that not only the collision energy but also system size may influence these trends, and further explorations are needed to unravel the physical phenomena determining the shape of the particle emitting source.

The Lévy-scale $R$ was also measured in Be+Be and Ar+Sc collisions at SPS, in Au+Au collisions at RHIC and in Pb+Pb collisions at the LHC. A characteristic dependence on transverse mass $m_T$ was found at every energy and centrality: $R$ decreases with $m_T$, as predicted based on the expanding nature of the fireball from which hadrons are created at the freeze-out.

Finally, correlation strength $\lambda$ was also measured in the above-mentioned systems. At RHIC, a decrease at small $m_T$ was found, and a saturation at larger $m_T$ values. The accessible range at LHC allowed only for the confirmation of the second trend. On the other hand, at SPS the mentioned decrease seemed to be absent, a difference that is expected to be explored in greater detail in higher precision SPS data as well.

The findings outlined above emphasise the importance of incorporating Lévy-shaped sources in femtoscopic analyses of momentum correlations measured in ultra-relativistic collisions of nuclei. To unravel the physical processes influencing the qualitative and quantitative details behind these observations, further phenomenological studies are required.

**Acknowledgments**

We acknowledge the support of NKFIH grants K-138136, TKP2021-NKTA-64, K-146913.

**References**

**Adams et al. (2005),** *Experimental and theoretical challenges in the search for the quark–gluon plasma: The STAR Collaboration's critical assessment of the evidence from RHIC collisions*, Nucl. Phys. A 757, 102 (2005).
**Adare et al. (2018)**, *Lévy-stable two-pion Bose-Einstein correlations in sqrt(sNN)=200 GeV Au++Au collisions*, Phys.Rev.C 97 (2018) 6, 064911
**Adcox et al. (2005),** *Formation of dense partonic matter in relativistic nucleus–nucleus collisions at RHIC: Experimental evaluation by the PHENIX Collaboration*, Nucl. Phys. A 757, 184 (2005).
**Adhikary et al. (2023)**, *Two-pion femtoscopic correlations in Be+Be collisions at sqrt(sNN)=16.84 GeV measured by the NA61/SHINE at CERN*, Eur.Phys.J.C 83 (2023) 10, 919
**Adler et al. (2007),** *Evidence for a long-range component in the pion emission source in Au + Au collisions at s(NN)\*\*(1/2) = 200-GeV*, Phys.Rev.Lett. 98 (2007) 132301
**Arsene et al. (2005),** *Quark–gluon plasma and color glass condensate at RHIC? The perspective from the BRAHMS experiment,* Nucl. Phys. A 757, 1 (2005).
**Back et al. (2005),** *The PHOBOS perspective on discoveries at RHIC*, Nucl. Phys. A 757, 28 (2005).
**Baym (1998)**, *The Physics of Hanbury Brown-Twiss intensity interferometry: From stars to nuclear collisions*, Acta Phys. Polon. B29, 1839 (1998)
**Csanád et al. (2007)**, *Anomalous diffusion of pions at RHIC*, Braz.J.Phys. 37 (2007) 1002-1013
**Csanád et al. (2024)**, *Femtoscopy with Lévy sources from SPS through RHIC to LHC*, Universe 2024, 10(2), 54
**Csörgő et al. (2004)**, *Bose-Einstein correlations for Levy stable source distributions*, Eur. Phys. J. C 36, 67 (2004)
**Csörgő et al. (2006)**, *Bose-Einstein or HBT correlation signature of a second order QCD phase transition*, AIP Conf. Proc. 2006, 828, 525–532.
**Glauber (1963)**, *Photon correlations*, Phys.Rev.Lett. 10 (1963) 84-86
**Goldhaber et al. (1959)**, *Pion-pion correlations in antiproton annihilation events*, Phys. Rev. Lett. 3, 181 (1959)
**Hanbury Brown et al. (1956)**, *A Test of a new type of stellar interferometer on Sirius*, Nature 178 (1956) 1046-1048
**Hanbury Brown (1991)**, *Boffin: A Personal Story of the Early Days of Radar, Radio Astronomy and Quantum Optics*, Bristol, Adam Hilger
**Kincses (2024)**, *Pion interferometry with Levy sources in sqrt(sNN)= 200 GeV Au+Au Collisions at STAR*, arXiv:2401.11169
**Lednicky (2001)**, Femtoscopy with unlike particles, in International Workshop on the Physics
**Lisa et al. (2005)**, *Femtoscopy in Relativistic Heavy Ion Collisions: Two Decades of Progress*, Ann. Rev. Nucl. Part. Sci. 55, 357 (2005)

**Nagy et al. (2018)**, *A novel method for calculating Bose–Einstein correlation functions with Coulomb final-state interaction*, Eur.Phys.J.C 83 (2023) 11, 1015
of the Quark Gluon Plasma (2001), arXiv:nucl-th/0112011
**Pórfy (2024),** *Femtoscopy analysis in small systems at NA61/SHINE*, arXiv:2401.02553
**Tumasyan et al. (2024)**, *Two-particle Bose-Einstein correlations and their Lévy parameters in PbPb collisions at sqrt(sNN) = 5.02 TeV*, Phys. Rev. C 2024 (accepted, in press), arXiv:2306.11574